# Project portfolio selection: Multi-criteria analysis and interactions between projects

Khadija BENAIJA[1], Laila KJIRI[2]

[1] ENSIAS, Université Mohammed-V, Rue Mohammed Ben Abdellah Regragui,
B.P. 713 Agdal, Madinat Al Irfane,
Rabat, Maroc.

[2] ENSIAS, Université Mohammed-V, Rue Mohammed Ben Abdellah Regragui,
B.P. 713 Agdal, Madinat Al Irfane,
Rabat, Maroc.

**Abstract**
In the project portfolio management, the project selection phase presents the greatest interest. In this article, we focus on this important phase by proposing a new method of projects selection consisting of several steps. We propose as a first step, a classification of projects based on the three most important criteria namely the value maximization, risk minimization and strategic alignment. The second step is building alternatives portfolio by the portfolio managers taking into account the classification of projects already completed in the first step. The third and final step enables the identification of the alternative portfolio to consider the contribution of projects to achieve the organization objectives as well as interactions between projects.

## 1. Introduction

In the 1980s, project management has seen an exceptional boom and many companies have adopted the principles and methods of project management. These methods (WBS, Gantt, PERT, logical framework ...) were directed to take projects individually.
Research will expand after its inquiries to cover all ongoing projects within an organization, and will be interested, in addition to efficient management of each project to the projects portfolio considered as a unit of global management.
Project Portfolio Management (PPM) provides answers to these questions:
• How to make sure that the projects will achieve the organization's strategic objectives?
• Are the limited resources (financial, human or material) allocated to good projects?
• Which projects should be selected to continue and which should be dismissed?

In this paper, we propose to examine the selection phase in the project portfolio management. The paper is structured as follows. In the next section, a literature review of projects portfolio management domain is provided, with an emphasis on the selection phase. Then, a method of multi-criteria analysis is presented enabling the classification of projects of a portfolio. The following section discusses a selection method based on the strategic value and interactions between projects. The last section includes our conclusions.

## 2. Literature review

Project portfolio management has received increasing attention in the last years, as the companies are launching more projects simultaneously.
A project portfolio is a collection of single projects and programs that are carried out under a single sponsorship and typically compete for scarce resources [1] [2].
This definition is similar to the one given by the Guide to the Project Management Body of Knowledge [3]: a portfolio is a collection of projects or programs and other works that are grouped together to facilitate effective management of that work to meet strategic business objectives.
The UK Office of Government Commerce (OGC) defines a portfolio as '…the totality of an organization's investment in the changes [projects and programs] required to achieve their strategic objectives' [4].

2.1 Project portfolio management

Portfolio management has become a priority for many companies enabling them to greatly improve their practice of project management in recent years. To be successful, a company must properly manage its projects. But first and foremost, we must manage good projects. It is the purpose of portfolio management: choosing the right projects.

Indeed, [5] states that the challenge for organizations is managing a potentially diverse range of projects while ensuring that the right projects are selected.

Markowitz was the first to introduce the concept of portfolio in the financial sector in 1952. His theory called Modern Portfolio Theory suggested that rational investors have to use diversification to optimize their portfolios: the portfolio in this case is a collection of financial assets and investments [6].

In 1981, McFarlan imported the portfolio management approach from the financial sector to the field of information technology (IT). He suggested that projects, rather than assets or investments, are the components of the portfolio and that the common management of projects - initially independent - may have advantages in achievement of business objectives of the company as well as reduction of the overall level of risk. [6]

The two complimentary drivers that led to the emergence of the concept of projects portfolio management are:
•   The need to make rational investment decisions that result in the delivery of organizational benefits [7].
•   The need to optimize the use of resources to ensure that the delivery of such benefits occurs in an effective and efficient manner [8].

Project portfolio management (PPM) includes the identification, prioritization, authorization, management and control of the component projects and programs and the associated risks, resources and priorities [6].

This concept enables:
•   New projects are evaluated and prioritized
•   Existing projects can be forestalled, cancelled or postponed
•   Resources are allocated and reallocated based on active projects

We can simplify this continuous process in four key steps as shown in Figure 1.

**Step 1: Identify and categorize projects**
This first step consists in taking inventory of projects (ongoing and potential). For each project inventoried, a project sheet is created. Once the inventory has been completed, we proceed to the classification by type of project, thus facilitating the subsequent steps. For example, it may be convenient to group the major projects vs minors, mandatory vs. discretionary, etc.

**Step 2: Evaluate and prioritize projects**
This step is very important; it is at this level that companies realize the biggest gain of the projects portfolio management namely the selection of the best projects. We present a review of the literature in more detail in the next section.

**Step 3: Authorize projects**
Having prioritized projects, this step aims to "draw the line" in determining which projects will be implemented. To complete this step, we carry out an analysis of the organizational capacity in order to maximize the use of available resources (human and financial). It is at the end of this stage that the project managers are assigned to different projects authority.

**Step 4: Report and revise portfolio**
This last step is to consolidate all the reports on the progress of various authorized projects. The goal is to give an overview to senior management, with dashboards that show the status and number of performance indicators. This information is crucial to facilitate decision-making by senior management on the continuation of projects.

The most important concepts of projects portfolio management discussed in the literature are summarized in table 1 below [6].

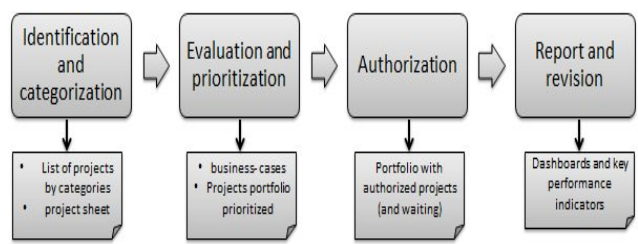

Fig. 1 Process for managing project portfolio.

Table 1: The concepts of PPM in literature [6]

| Article | Project identification, categorisation and prioritisation | Project opportunity assessment, selection and portfolio balancing | Portfolio performance management and review | Portfolio governance | Portfolio resource management | Portfolio communication and change management | Portfolio risk management |
|---|---|---|---|---|---|---|---|
| PMI, 2008[9][3] | x | x | x | x | | | x |
| OGC, 2011[10] | x | x | x | x | x | x | x |
| OGC, 2007[11] | | | x | x | x | x | x |
| PM Solutions, 2007[12][13] | x | x | x | x | x | x | |
| Parviz, and Levin, 2006[14] | x | x | x | x | | x | |
| APM, 2006[15] | | | | x | | | |
| Krebs, 2009[16] | x | x | | | x | | |
| IPMA, 2008[17] | x | | | | | | |
| Artto and Dietrich, 2004[18] | x | x | x | x | | x | |
| Archer and Ghasemzadeh, 1999[1] | x | x | x | | | | |
| Blichfeldt and Eskerod, 2005[19] | x | | | | | | |
| Blomquist and Muller, 2006[20] | x | x | | | | | x |
| Bouraad, 2008[21] | | | | x | | | |
| Cooper and al., 2001[22] | x | x | x | | x | | x |
| Engwall and Jerbrandt, 2002[23] | | | | | x | | |
| Iamratanakul and Milosevic, 2007[24] | x | x | x | | | | |
| Kendall and Rollins, 2003[25] | x | x | x | | x | | |
| Killen et al., 2008[26] | | x | | | | | |
| Levine, 2005[27] | x | x | x | | x | | |
| Patanakul and Milosevic, 2005[28] | | | x | x | | | |
| Petit and Hobbs, 2010[29] | | | | | | | x |
| Holland and Fathi, 2007[30] | | | | | | | x |
| Meskendahl, 2010[31] | | x | | | | | |

We limit ourselves in this article to study the concept of projects evaluation and prioritization. The process of projects and programs selection is considered to be the main component of the portfolio management system [32].

2.2 Evaluation and prioritization of projects

As we saw in the previous section, the key to success in managing a portfolio of projects is to choose the right projects at the right time [27]. Let's start by defining the evaluation and prioritization of projects.
Evaluate projects: This step aims to document the projects in order to compare them. Evaluation is ultimately to build a business case that establishes the costs and deadlines of project, benefits, advantages / disadvantages, risks, etc... The business case allows having a common basis for evaluation of projects.
Prioritize projects: At this stage, projects are compared in order to determine priorities. The use of multi-criteria matrix (scoring models) is recognized as a best practice.
We are then facing a multi-criteria problem. Indeed, it is a decision-making to build a portfolio of projects that best achieve the organization's strategic objectives [32].
From an initial situation defined by a set of identified projects (ongoing and candidates), and a set of criteria based on the objectives of the organization, it must propose alternatives portfolio and evaluate them in order to determine the best one. We are led to consider the following sets:
- The list of projects is obtained from stage 1 (see previous section)
- The set of alternatives: In organizations implementing many projects at the same time, the number of alternative portfolios can be very large, and as a result the problem may be of combinatorial nature [32].
- The set of criteria
- The set of evaluations of alternatives according to selected criteria.

We can define the notion of criteria as a tool for measuring the degree of success of a particular objective [32].
The determination of the criteria is an important task that requires detailed analysis. According to [33], this choice must essentially satisfy these three conditions:
- Completeness: The criteria should reflect all important aspects.
- Consistency: it is maintained if local relationships between portfolio alternatives (each criterion taken separately) are consistent with the relationship at the global level (with respect to all criteria).
- Lack of redundancy: It means that the concepts measured in a criterion are not repeated in another. In other words, the removal of a criterion leads to a dissatisfaction of at least one of the other conditions.

The criterion can be quantitative or qualitative. It is easier to measure a quantitative criterion; in this case, an already agreed scale is used. For example, we can use a monetary scale to evaluate Net Present value criterion. For qualitative criterion, we have to use a subjective scale since an objective scale usually does not exist [32].
Let us study the criteria for evaluating project portfolios proposed in the literature.
The Standard for Portfolio Management published by PMI [9] proposes a classification of criteria used for portfolio evaluation. This classification suggests that the analysis of a projects portfolio should cover the following aspects: general business criteria, financial criteria, risk related criteria, criteria for evaluating the project's compliance with the requirements of the current legal situation, criteria for analyzing human resource management issues, marketing criteria, and technical criteria.
We find approximately the same areas in the proposal of Meredith and Mantel [34]. These suggest that the criteria should allow the evaluation of projects in the following areas: production, marketing, finances, staff, administration and other categories. In addition, they propose several evaluation criteria for each domain.
The selection of the criteria should be determined by the specificity of the organization [32]. Nevertheless, we can highlight some key objectives that a project portfolio must achieve. Through the review of literature, the most important objectives are:
- Maximization of organization's value;
- Balancing the portfolio (in order to minimize the risk);
- Adjusting portfolio to organization's strategy (strategic alignment).

The approach adopted in this paper is represented in Figure 2.

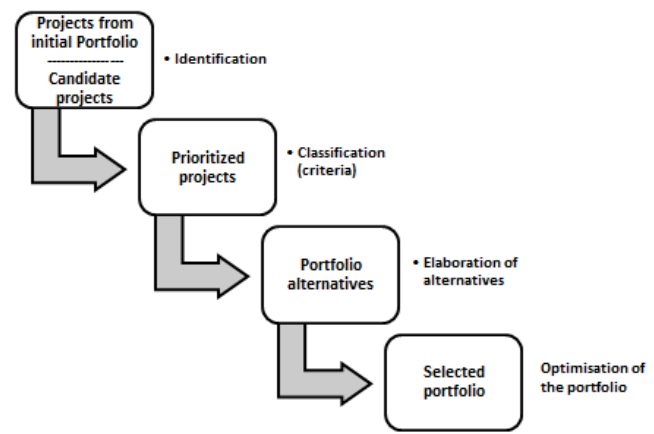

Fig. 2 The approach adopted for portfolio selection in this paper.

This approach can be summarized as follows:
**Initial state:**
We have two lists of projects: the list of all projects in the initial portfolio and the list of candidate projects (identified as potential).
**Step 1: classification of projects of the two lists**
By applying the classification proposed in Section 3 according to three criteria: value maximization, risk minimization and strategic alignment, we get a list of prioritized projects.
**Step 2: Elaboration of portfolio alternatives**
From the list of prioritized projects, the portfolio managers can form portfolio alternatives. A portfolio alternative may include existing projects in the initial portfolio, and candidate projects. We are led to manage one of the following cases:
*Case 1:* Adding a new project to the portfolio.
*Case 2:* removing a project from the portfolio.
**Step 3: Selection of portfolio alternative**
The method presented in Section 4 allows managing both cases above through the calculation of the strategic value of the project and the overall cost and time. These three parameters are used to facilitate decision-making on the selection of portfolio alternative.

## 3. Classification framework based on a three-dimensional analysis

As we saw in the previous section, several criteria exist in the selection of projects. In the framework that we present, we propose to retain the three most important criteria: value, risk and strategic alignment. Note that these criteria can be changed depending on the choice of project portfolio managers.

This classification framework results from a previous work [35].

### 3.1 The bivariate analyses

Considering the three criteria mentioned below, we can make the following bivariate analyses: value / risk analysis, risk / alignment analysis and value / alignment analysis.

**Analysis risk-value:**
Let us consider the two-dimensional analysis risk-value, as presented by [36].

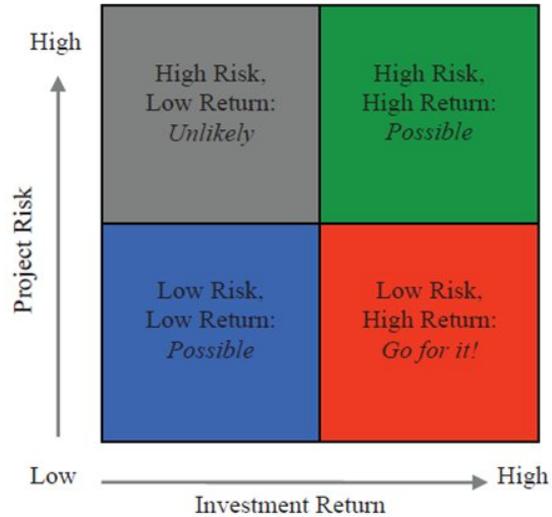

Fig. 3 Central idea of diversifying the portfolio: Managing risk and value creation [36].

As shown in figure 3, we can identify high-potential projects according to these two criteria: risk and value.
  * Projects with low risk and high value are preferred: they have great potential because they generate more value with low uncertainty.
  * Projects with a high risk and low value are discarded. They are the opposite of the first ones: they generate a low value with a large uncertainty.
  * Projects with a high risk and a high value as well as those with low risk and low value must be managed according to company strategy. Hence, it is necessary to call the third criteria: strategic alignment.

**Analysis risk-alignment:**
We can perform a similar analysis by considering two criteria: risk and strategic alignment. We can classify the projects in the portfolio (or just candidates) under three headings:
  * Projects with low risk and very aligned with the business strategy: these projects are to retain.
  * Projects with a high risk and non-aligned with the business strategy: these projects must be discarded.
  * Projects with a high risk and very aligned with the business strategy or those with a low risk but not aligned with the strategy: these projects should be subject to manager's decision, taking into account one or more criteria, including the value generated by these projects.

**Analysis value-alignment:**
The third two dimensional analysis concerns two criteria: value and strategic alignment. We propose, as for earlier analysis, this classification of projects into three categories:

* Projects with a high value and very aligned with the strategy, these projects must be selected;
* Projects with a low value and not aligned with the strategy: these projects must be discarded;
* Projects with a high value but not aligned with the strategy or those with a very low value and aligned with strategy: these projects require a decision.

3.2 Trade-off between the three criteria

After bivariate analysis, we propose to combine the three criteria by considering a three-dimensional analysis.

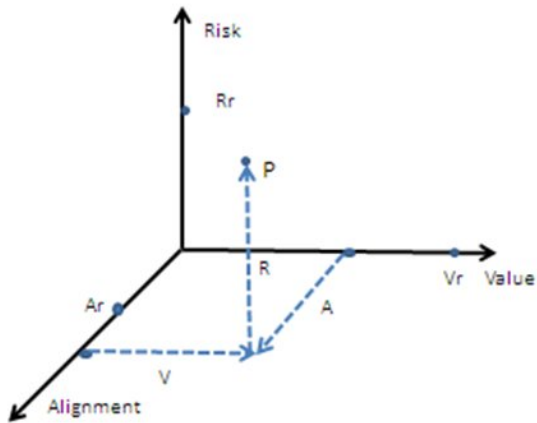

Fig. 3 The three coordinates of a project [35].

If we consider a project P, we can estimate its level of risk R, its expected value V and its level of alignment with the business strategy A. These three coordinates are placed in a three-dimensional reference frame, as shown in figure 4. These values are estimated using methods that are not discussed in this paper.

Rr is the risk value from which we can say that the risk is high and below which the risk is considered tolerable.
Vr is the value from which the benefit is considered important, and below which it is considered low.
Ar is the value of the alignment from which it is considered high, and below which it considered low.
We attribute "+" if the value is better than the reference value and "-" if it is worse. We can translate it into:
* + for R < Rr or V >= Vr or A >= Ar
* - for R >= Rr or V < Vr or A < Ar

Table2: Scoring of possible cases [35].

| Case | Coordinates | Scoring |
|---|---|---|
| Case 1 | V >= Vr and R < Rr and A >= Ar | + + + |
| Case 2 | V >= Vr and R < Rr and A < Ar | + + - |
| Case 3 | V >= Vr and R >= Rr and A < Ar | + - - |
| Case 4 | V >= Vr and R >= Rr and A >= Ar | + - + |
| Case 5 | V < Vr and R < Rr and A >= Ar | - + + |
| Case 6 | V < Vr and R < Rr and A < Ar | - + - |
| Case 7 | V < Vr and R >= Rr and A >= Ar | - - + |
| Case 8 | V < Vr and R >= Rr and A < Ar | - - - |

For a given project, one of the following cases occurs:
**Case 1:** V >= Vr and R < Rr and A >= Ar: This case will be appreciated "+ + +".
**Case 2:** V >= Vr and R < Rr and A < Ar: This case will be appreciated "+ + -".
**Case 3:** V >= Vr and R >= Rr and A < Ar: This case will be appreciated "+ - -".
**Case 4:** V >= Vr and R >= Rr and A >= Ar: This case will be appreciated "+ - +".
**Case 5:** V < Vr and R < Rr and A >= Ar: this case will be appreciated "- + +".
**Case 6:** V < Vr and R < Rr and A < Ar: this case will be appreciated "- + -".
**Case 7:** V < Vr and R >= Rr and A >= Ar: this case will be appreciated "- - +".
**Case 8:** V < Vr and R >= Rr and A < Ar: this case will be appreciated "- - -".

By analysing the eight cases, we can refine the classification in four categories according to the degree of potential, materialized by the number of "+":

Table3: Classification framework [35].

| Rubric | Degree of potential | corresponding case | Decision |
|---|---|---|---|
| Rubric 1 | 3+ | Case 1 | to select |
| Rubric 2 | 2+ | case 2, case 4, case 5 | to prioritize |
| Rubric 3 | 1+ | case 3, case 6, case 7 | to lower priority |
| Rubric 4 | 0+ | Case 8 | to abandon |

**Rubric 1:** including the case 1 with a rating of three "+", projects of this category must be selected.
**Rubric 2:** including cases 2, 4, and 5 with a rating of two "+", projects in this section are interesting to select. For example, if the company gives more priority to the creation of value and risk management it must choose projects of case 2 (for risk and alignment the case 5, and for value and alignment: case 4).
**Rubric 3:** including cases 3, 6 and 7 with a rating of one "+", the projects in this section are low potential.
**Rubric 4:** including the case 8, projects of this section are to give up as all criteria are negative.

Thus, the existing projects in the portfolio or candidates to be selected can be classified into these four categories.
The criteria evaluation was applied at the project level and not at the portfolio level. Let us remember that we are at the stage "Evaluation and prioritization of projects" of Figure 1, and the purpose of the framework we have presented in this section is precisely to propose a classification by priority of identified projects.

Once this classification is made, we can refine the selection of projects by considering the contribution of each project to the achievement of the business objectives, and also considering the interactions between projects. This is what we will treat in the next section.

## 4. Strategic value and interactions between projects

The problem of selection of project portfolio is a continuous process, it consists generally to answer this question: what are the new proposals to be included in the portfolio and what are the projects that should be removed? [32].

To answer this question, we propose to follow a two-step approach. The first one is the classification of projects according to specific criteria: the classification presented in the previous section. The second step is to evaluate the introduction of a project - identified as high potential project in the classification - in the portfolio. This step has been the subject of previous work [37].

We are led to manage this situation: portfolio includes N projects already launched: $P_1, \ldots P_N$, in addition, we have candidate projects.

As a first step, we present an approach to include the project P to the portfolio, taking into account the maximization of portfolio contribution in achieving business objectives. Then, we study the introduction of project P in the portfolio, considering the optimization of interactions with other projects.

4.1 Maximization of the strategic value

Strategic planning of the organization is implemented through the (or) portfolio (s) of projects in order to achieve objectives. The projects are then the means for implementing this strategy.

The strategic benefits are therefore a link between projects and portfolio objectives. Each project brings new skills, new knowledge or improvements to the organization.

A cancelled project can bring benefits even if it has been stopped before its end. The final deliverable is not binding on the project to the objectives; it is the benefit of deliverable that binds the project to the objectives [38].

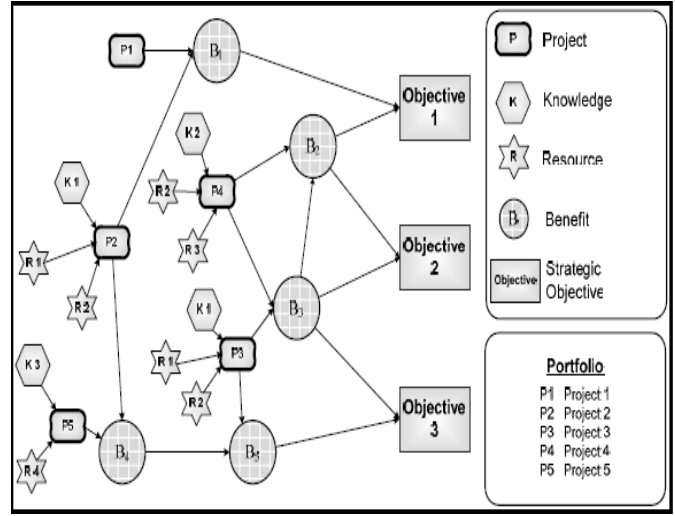

Fig. 5 The Interdependence Model [38].

As the interdependencies model above shows, the project interdependences are not defined through strategic elements falling in a cascade down to projects as if they were independent, but a network where projects are interconnected in terms of their contribution to the benefits and final goals of the organization [38].

It is meant by "resource" all strategic resources such as specialists, scientists or specific equipment, those present inputs to the projects. Non-strategic resources are usually not shown in this model in order to simplify its representation. To simplify the modelling of portfolios consisting of a large number of projects, we can group multiple projects that share the same strategic resources and knowledge into a single project. Also, the knowledge developed by a project may be used as input for another one [38].

The assessment of the project contribution to the achievement of a portfolio objective $O_j$ is expressed as follows:

$$C_{POj} = \sum_{i=1}^{B} ( C_{PBi} * C_{BiOj} )$$

(1)

Where

$C_{POj}$: Contribution of project P to objective $O_j$.
$C_{PBi}$: Relative contribution of project P to key benefit $B_i$.
$C_{BiOj}$ : Relative contribution of key benefit $B_i$ to objective $O_j$.
P : Project in the stream project-benefit-objective.
$O_j$: Objective in the stream project-benefit-objective.
$B_i$: Key benefit in the stream project-benefit-objective.
B: Total number of benefits which the project P contributes, those profits that contribute the objective $O_j$.
This formula is adapted from [38].

We can then propose the calculation of the strategic value of project P $V_{sp}$ as follows:

$$V_{sp} = \sum_{j=1}^{J} C_{POj} \qquad (2)$$

Where J is the total number of objectives which the project P contributes.

We can even introduce a weighting according to the importance of the objective in the company strategy as follows:

$$V_{sp} = \sum_{j=1}^{J} a_j * C_{POj} \qquad (3)$$

Where $a_j$ represents the weighting coefficient of the objective $O_j$.

The strategic value $V_{sp}$ of project P will be made available to decision makers within the company or portfolio managers as a criterion for introducing the project P to the portfolio. On the other hand, it can be a criterion for removing an existing project in the portfolio.

### 4.2 Optimization of interactions between projects

Let us consider the portfolio with N projects ($P_1$, … $P_N$). We propose to study the impact of the introduction of the project P on this portfolio, according to the two important elements, namely: the cost and time frame.

Considering the approach as a result, we suppose that the project P is introduced and we evaluate the new costs and delays of ongoing projects.

$C'_{Pi}$ is the new estimated cost of the project $P_i$, $C_{Pi}$ is his evaluated cost before the introduction of the project P:

$$C'_{Pi} = a_i * C_{Pi} \qquad (4)$$

Where $a_i$ is a coefficient $\geq 0$,
$a_i = 0$ if P include $P_i$,
$0 < a_i < 1$ if P reduce the $P_i$ cost,
$a_i = 1$ if P does not impact the $P_i$ cost,
$a_i > 1$ if P increase the $P_i$ cost.

Similarly, we consider the impact on time frame as follows. $D'_{Pi}$ is the new estimated completion time of project $P_i$ and $D_{pi}$ its estimated completion time before the introduction of project P to the portfolio.

$$D'_{Pi} = b_i * D_{Pi} \qquad (5)$$

Where $b_i$ is a coefficient $\geq 0$,
$b_i = 0$ if $P_i$ will be stopped before start,
$0 < b_i < 1$ if P will reduce the $P_i$ time frame,
$b_i = 1$ if P does not impact the $P_i$ time frame,
$b_i > 1$ if P will delay the completion of $P_i$.

We can then calculate the overall additional cost of the portfolio as follows:

$$C_G = \sum_{i=1}^{N} (C'_{Pi} - C_{pi}) \qquad (6)$$

We can introduce, here too, weighting according to the criticality and sensitivity of projects:

$$C_G = \sum_{i=1}^{N} k_i * (C'_{pi} - C_{pi}) \qquad (7)$$

Where $k_i$ is a cost sensitivity coefficient of $P_i$ in the portfolio.

Similarly, we define the overall impact on time frames of the portfolio as follows:

$$D_G = \sum_{i=1}^{N} (D'_{pi} - D_{pi}) \qquad (8)$$

And taking into account the weighting:

$$D_G = \sum_{i=1}^{N} l_i * (D'_{pi} - D_{pi}) \qquad (9)$$

Where $l_i$ is a time frame sensitivity coefficient of $P_i$ in the portfolio.

The $C_G$ and $D_G$ values can be positive or negative, depending on the impacts of P on the portfolio.

The calculation of three parameters $V_{sp}$, $C_G$ and $D_G$ allows decision makers to choose the most advantageous portfolio alternative.

## 5. Conclusions

Effective management of the projects portfolio is one of keys to success of any organization. Indeed, it is not sufficient to clearly define the objectives to be attained but we must choose the right means to achieve them. These means are none other than the projects.

The method of projects selection described in this article is part of the projects portfolio management. It is based on the most important criteria that emerge from the review of the literature in this field. It uses an interactive approach due to the intervention of decision makers throughout the process. For example, we do not analyze all the possible portfolio alternatives that can be very numerous, but it is the managers who define the alternatives that need to be studied on the basis of the classification they have.

Checking these elements, the implementation of this method is required; this is one of the perspectives of this work.

**Khadija BENAIJA** Diploma of computer engineer in 1994 from the Ecole Mohammadia d'Ingénieurs (EMI) at the University Mohammed V, Rabat, Morocco. Currently, a PhD student since 2013 at Alqualsadi research and development team, ENSIAS, University Mohammed V, Rabat, Morocco. Three papers published so far in the field of projects portfolio management. Program Manager since 2003 at CIO of Telecommunications company, Morocco.

**Laila KJIRI** 3rd cycle doctorate in Computer Science in 1986 at the University GrenobleII, Grenoble, France. Ph-D Computer in 1995 at the University of Montreal, Quebec, Canada. Currently Teaching and research in Engineering Software Department, since 1995 at ENSIAS, University Mohammed V, Rabat, Morocco.